\documentclass[conference]{IEEEtran}

\usepackage{amsthm}

\IEEEoverridecommandlockouts

\usepackage{bm,cite,algorithm,algorithmic,float,amsmath,amssymb}
\usepackage[dvips]{graphicx}
\usepackage{subfigure,amsfonts}
\usepackage{cite, color}
\usepackage{balance}
\floatname{algorithm}{Algorithm}
\ifCLASSINFOpdf
\else
\fi
\begin{document}
%
\title{Amplify-and-Forward Full-Duplex Relay with Power Splitting-Based SWIPT}

\author{\IEEEauthorblockN{Hongwu Liu}
\IEEEauthorblockA{Shandong Jiaotong University\\
Jinan, China\\
Email: hong.w.liu@hotmail.com}
\and
\IEEEauthorblockN{Kyung Sup Kwak}
\IEEEauthorblockA{TelLab, Inha University\\
Incheon, Korea\\
Email: kskwak@inha.ac.kr}}

\maketitle

\begin{abstract}
This paper proposes a virtual harvest-transmit model and a harvest-transmit-store model for amplify-and-forward full-duplex relay (FDR) networks with power splitting-based simultaneous wireless information and power transfer. The relay node employs a battery group consisting of two rechargeable batteries. By switching periodically between two batteries for charging and discharging in two consecutive time slots of each transmission block, all the harvested energy in each block has been applied for full duplex transmission in the virtual harvest-transmit model. By employing energy scheduling, the relay node switches among the harvesting, relaying, harvesting-relaying, and idle behaviors at a block level, so that a part of the harvested energy in a block can be scheduled for future usage in the harvest-transmit-store model. A greedy switching policy is designed to implement the harvest-transmit-store model, where the FDR node transmits when its residual energy ensures decoding at the destination. Numerical results verify the outage performance of the proposed schemes.
\end{abstract}

\IEEEpeerreviewmaketitle

\section{Introduction}

Energy harvesting (EH) has emerged as a promising enabling technology for wireless relaying networks \cite{Energy_harvesting_relay, SWIPT_protocol_AF}. By harvesting energy from ambient radio-frequency (RF) signals, periodic battery replacement or recharging can be alleviated for energy-constrained relay nodes. Since RF signals can carry both information and energy, simultaneous wireless information and power transfer (SWIPT) has gained a lot of interests from academic institutions and industry \cite{Grover_Shannon_meets_Tesla, SWIPT_cellular,SWIPT_magazine,SWIPT_protocol_AF}. Recently, two practical receiver architectures, namely, time switching (TS) and power splitting (PS) \cite{SWIPT_architecture}, have been adopted in various SWIPT systems \cite{SWIPT_cellular, SWIPT_OFDM, MIMO_B_SWIPT}.
In TS-based SWIPT (TS-SWIPT), the receiver harvests power from an
energy signal sent by the source and then receives the source transmitted information signal in
a time-division manner. In PS-based SWIPT (PS-SWIPT), the receiver extracts energy from
the received source signal with the aid of PS. In general, PS-SWIPT reduces the time slots
consumed compared with TS-SWIPT, so that the information transmission time, as well as
the spectral efficiency, can be increased.

By employing TS-based and PS-based protocols for amplify-and-forward (AF) relay networks \cite{SWIPT_protocol_AF},  SWIPT not only keeps energy-constrained relay nodes active, but also enables information relaying across barriers or over long distance. In \cite{SWIPT_SRR}, the outage and diversity performances of SWIPT in cooperative networks with spatially random relays have been investigated. In \cite{SWIPT_DPS}, the distributed PS-based SWIPT has been designed for interference-limited relay networks. Several power allocation schemes for EH relay networks with multiple source-destination pairs were investigated in \cite{SWIPT_PA}. Furthermore, antenna switching and antenna selection have also been applied for SWIPT relaying networks \cite{SWIPT_antenna_switch, SWIPT_Antenna_select}. In \cite{SWIPT_AF_DPS2}, PS-SWIPT has been investigated for AF relaying networks by employing full and partial channel state information. All the aforementioned relay-assisted SWIPT employ half-duplex relay (HDR) nodes, so that two time phases are needed to accomplish one time of transmission.

Since full-duplex relay (FDR) can receive and transmit simultaneously, the spectral efficiency of an FDR network can be significantly improved over its HDR counterpart. Recently, the applications of SWIPT in FDR networks have drawn much attention \cite{SWIPT_FD_selfenergy, SWIPT_FD, SWIPT_FDR_MIMO}.
By employing separated relay receive antenna and transmit antenna for EH and information relaying, respectively, the authors of \cite{SWIPT_FD_selfenergy} proposed a self-interference immunizing FDR scheme.
In \cite{SWIPT_FD}, the throughput has been analyzed for FDR networks with TS-SWIPT. Then, MIMO antennas have been employed at the relay to enhance the performance of TS-SWIPT in FDR networks \cite{SWIPT_FDR_MIMO}. Due to TS implementation, all the aforementioned TS-SWIPT schemes in FDR networks are not strictly operated in FDR mode, so that the reduction of information transmission time is unavoidable. On the other hand, PS-SWIPT has shown its performance improvement over TS-SWIPT in HDR networks \cite{SWIPT_protocol_AF}. Since PS-SWIPT does not change the effective information transmission time in relay networks \cite{SWIPT_protocol_AF}, it is suitable to employ PS-SWIPT in FDR networks, so that the effective information transmission time can be doubled compared to that of HDR networks.

To the best of our knowledge, how to deploy PS-SWIPT in FDR networks is still an open problem.
The technical challenge of implementing PS-SWIPT in FDR network is how to realize full-duplex
energy harvesting and information relaying, i.e., charging and discharging simultaneously
at the relay node besides the full-duplex information processing. In this paper, we propose to employ a battery group consisting of two batteries at the relay node to realize the full-duplex operation.
By periodically switching between two rechargeable batteries for charging and discharging during two consecutive time slots of each block, the energy-constrained relay can be self-powered in a virtual harvest-transmit model. The harvest-transmit-store model along with its greedy switching (GS) policy has also been designed.

The rest of this paper is organized as follows. Section II describes the system model and
the virtual harvest-transmit model of the considered AF FDR-assisted PS-SWIPT. Section III presents
the harvest-transmit-store model and proposes the GS policy for its implementation. Section IV presents
numerical results and discusses the system performance of our proposed scheme. Section
V summarizes this study.

\section{System Model}

In the considered wireless FDR network, a source intends to transmit its information to a destination. Due to physical isolation between the source and destination, an AF FDR node is employed to realize the dual-hop relay transmission. The source and destination are equipped with a single antenna, respectively, whereas the relay node is equipped with a single receive antenna and a single transmit antenna. All the channels are assumed to be quasi-static block fading, i.e., the channel coefficients keeps constant during one block and vary independently from block to block.
The channels of the source-to-relay and relay-to-destination links are denoted by $h_1 = \sqrt{{\mathcal{L}_1}} \tilde h_1$ and $h_2 = \sqrt{{\mathcal{L}_2}} \tilde h_2$, respectively, where ${\mathcal{L}_i}$ and $\tilde h_i$ ($i=1, 2$) are the large-scale path-loss and small-scale fading of two-hop links, respectively.
For the sake of exposition, the channel gain of $h_i$ is denoted by $g_i \triangleq |h_i|^2$  for $i \in \{1,2 \}$.
The small-scale channel magnitude, $|\tilde h_i|$ ($i=1, 2$), is modeled as Nakagami-$m$ fading with the unit mean such that $g_i$ ($i=1, 2$) is distributed according to the gamma distribution with the shape factor ${m_i} $ and the scale factor $\theta _i \triangleq \tfrac{\bar g_i}{m_i} $. Further, the normalized transmitted signals of the source and relay are denoted by $x_s(t)$ and $x_r(t)$, respectively. The transmission powers at the source and relay are denoted by $p_s$ and $p_r$, respectively.

In order to enable charging and discharging simultaneously, the energy-constrained relay deploys a battery group consisting of two rechargeable batteries, as depicted in Fig. 1. The two batteries are assumed having the same initial state. The duration of each block, $T$, is divided equally into two time slots (odd and even slots). The two batteries are activated for EH and power supplying alternately in the odd and even slots during each block. In the odd (even) slot of a block, battery \#1 (battery \#2) functions in discharging, while battery \#2 (battery \#1) switches to the EH receiver for charging. Further, in each block, the consumed energy quantum of battery \#1 (battery \#2) in the odd (even) slot is set to equal to that of the relay-harvested energy during the even (odd) slot. Following the above procedures, the full-duplex relaying is powered in a self-sustainable way. Since this charging/discharging  behavior  mimics the harvest-transmit model of PS-SWIPT in HDR networks \cite{SWIPT_protocol_AF}, where a single battery has been applied, we call it the virtual harvest-transmit model.
With the aid of channel estimation designed for energy-constrained networks \cite{Optimal_Training_WET}, we assume that the relay has the capability to access full channel state information.

\begin{figure}[htbp]
\begin{center}
\includegraphics[width=2in]{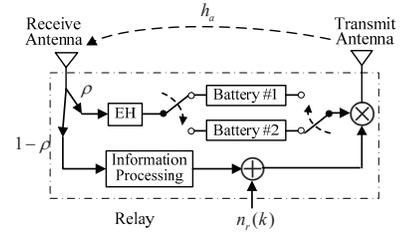}
\vspace{-0.09in}
\caption{Block diagram of the AF FDR node with PS operation.}
\end{center}
\vspace{-0.25in}
\end{figure}

In each slot, the incident signal at the relay receive antenna can be expressed as
\begin{eqnarray}
y_r(t) &\!\!\!\!=\!\!\!\!&  \sqrt{p_s} h_1 x_s(t) + \sqrt{p_r} h_a x_r(t) + n_a (t),
\end{eqnarray}
where $h_a$ is the residual self-interference (RSI) channel incident on the receive antenna and $n_a(t) \sim \mathcal{CN}(0, \sigma_a^2)$ is the additive noise at the receive antenna.
The power of the relay-received signal is split in $\rho : 1- \rho$ proportion for EH and information processing, where $\rho$ is the power splitting ratio.
Due to a negligible power of antenna noise, the harvested energy at the end of a time slot can be written as
\begin{eqnarray}
{E_h} =\eta  \rho \left( {  {p_s}g_{1}}  +  p_r g_{a} \right)\tfrac{T}{2}, \label{eq:e_h}
\end{eqnarray}
where $\eta $ ($0< \eta  <1$) is the energy
conversion efficiency and $g_a \triangleq |h_a|^2$ is the RSI channel incident on the relay receive antenna. Note that $E_h$ has been sent to battery \#2 (battery \#1) for charging in the odd (even) slot. Simultaneously, the relay transmission in the current slot is powered by the battery that is not switched for EH. When $E_h$ is adopted as the transmission energy quantum, the relay transmission power can be characterized by
\begin{eqnarray}
p_{r} = \frac{{{E_h}}}{{T/2}} = \eta  \rho \left( {  {p_s}g_{1}}  +  p_r g_{a} \right). \label{eq:P_max}
\end{eqnarray}
Based on \eqref{eq:P_max}, the relay transmission power can be expressed as
\begin{eqnarray}
p_{r} = \frac{\eta\rho p_s g_1}{1-\eta\rho g_a}.
\end{eqnarray}

The sampled signal at the relay for information processing can be expressed as
\begin{eqnarray}
\!\!\!\!\!\!y_r(k) &\!\!\!\!\!\!=\!\!\!\!\!\!&  \sqrt{(1\!-\!\rho)p_s} h_1 x_s(k) \!+\! \sqrt{(1\!-\!\rho)p_r} h_b x_r(k) \!+\! n_r (k), \label{eq:r_signal}
\end{eqnarray}
where $k$ denotes the symbol index, $h_b$ is the RSI channel remained in the digital-domain after some stages of interference cancellation, $n_r(k) \triangleq \sqrt{1\!-\!\rho} n_a(k) + n_p(k)$ is the additive noise with the zero mean, $n_a(k)$ is the sampled version of the antenna noise $n_a(t)\mathcal{CN}(0, \sigma_a^2)$, $n_p(k)\sim \mathcal{CN}(0, \sigma_p^2)$ is the processing noise. Since $n_r(k)$ is dominated by the processing noise rather than the antenna noise, we approximate that $n_r(k)$ has the variance $\sigma_r^2 \approx \sigma_p^2$.
The signal $x_r(k)$ in \eqref{eq:r_signal} is expressed as $x_r(k) = G y_r(k-\tau)$, where $G=1/\sqrt{(1-\rho)p_s g_1 + (1-\rho) p_r g_b + \sigma_r^2 }$ is the amplification coefficient and $\tau$ is an amount of delayed symbols due to signal processing at the relay. The received signal at the destination is given by
\begin{eqnarray}
y_d(k) = \sqrt{p_r} h_2 x_r(k) + n_d (k),  \label{eq:d_signal}
\end{eqnarray}
where $n_d(k)\sim \mathcal{CN}(0, \sigma_d^2) $ is the noise at the destination. For this system, the instantaneous end-to-end (e2e) signal to interference plus noise ratio (SINR) is expressed as
\begin{eqnarray}
\!\!\!\!\gamma_{\rm e2e} &\!\!=\!\!& \frac{\gamma_r \gamma_d}{\gamma_r + \gamma_d +1} \nonumber \\
\!\!\!\! &\!\! \approx \!\!& \min\{ \gamma_r, \gamma_d \} {\rm ~~for~intermediate/high~SINRs},  \label{eq:gamma_e2e}
\end{eqnarray}
where $\gamma_r \triangleq \tfrac{(1-\rho)p_s g_1}{(1-\rho)p_r g_b + \sigma_r^2} $, $\gamma_d \triangleq \tfrac{p_r g_2}{\sigma_d^2}$, and $g_b \triangleq |h_b|^2$ is the RSI channel gain in the digital-domain after some stages of interference cancellation.

\section{PS-SWIPT with Energy Scheduling}

In order to decode the relaying signal received at the destination, it requires that the e2e SINR at least equals to a target value $\gamma_{\rm th}$. In the virtual harvest-transmit model, the harvested energy in each block is directly used for relay transmission, without considering energy scheduling across channel realizations. Although the virtual harvest-transmit model is easy to implement, it would perform better if energy scheduling is allowed to store a part of the harvested energy for future usage. In this section, we propose the harvest-transmit-store model with its GS implementation for the considered network.

Based on PS operation and time-switching between two batteries, the relay can simultaneously charge one battery and forward its received signal with the stored energy of another battery. According to the channel condition, the relay can also perform only EH or relaying. Thus, the relay can switch among four operational modes: a) $\mu_h$: the two batteries harvest energy from the relay-received signal during the odd and even slots, respectively, b) $\mu _r$: the relay transmits data with its power being supplied by the two batteries in the odd and even slots, respectively,  c) $\mu _{hr}$: the relay harvests energy and forward data as that of the virtual harvest-transmit model, and d) $\mu _{\phi}$: the relay neither harvests nor transmits when both EH and transmission are impossible. In the $t$-th  block, the operational mode of the relay is denoted by $\mu (t) \in \{\mu _h, \mu_r, \mu _{hr}, \mu_{\phi} \} $.
Since the decoding at the destination depends on whether the relay transmits or not, the relay transmission power can be expressed as
\begin{eqnarray}
p_r \!=\! \left\{ {\begin{array}{*{20}{c}}
\!\!  \tfrac{\gamma_{\rm th} \sigma_d^2}{g_2},  &\!\!\!\!\!\!\!\!   \tfrac{ p_s g_1}{\sigma_r^2} > C_1 {\rm ~and~} \mu(t)=\mu_{hr} \\
\!\!  \tfrac{\gamma_{\rm th} \sigma_d^2}{g_2},  &\!\!\!\!\!\!\!\!  \gamma_{\rm th}  < \tfrac{ p_s g_1}{\sigma_r^2} \le C_1 {\rm ~and~} \mu(t)=\mu_{r}\\\
\!\!  {\rm does~not~exist,} & \!\!\!\!\!\!\!\!\!\!\!\! {\rm otherwise}
\end{array}} \right.\!\!\!\! ,  \label{eq:p_r_acc}
\end{eqnarray}
where $C_1=\tfrac{\gamma_{\rm th}((1-\rho)g_b \gamma_{\rm th} \sigma_d^2+g_2\sigma_r^2)}{(1-\rho)\sigma_r^2}$ is obtained by substituting $p_r = \tfrac{\gamma_{\rm th} \sigma_d^2}{g_2}$ into $ \gamma_{r} = \gamma_{\rm th}$.

We assume that the two batteries at the relay have the same size $p_b = \alpha p_s$ with $\alpha>0$. Each battery is discretized into $L+2$ energy levels $\varepsilon _i \triangleq i p_b/(L+1)$, where $i=0, 1, \ldots, L+1$ \cite{Lifetime_max, Relaying_vs_EH}. We define $s_i$, $i=0, 1, \ldots, L+1$  as $L+2$ energy states for each battery, so that each battery is in state $x_i$ when its stored energy equals to $\varepsilon _i$. Further, $P_{i,j}$ denotes the transition probability $\Pr\{s_i \to s_j \}$ and $E_0(t) \in \{ \varepsilon _i : 0 \le i \le L+1   \}$ denotes the residual energy of each battery at the beginning of the $t$-th block. Also, we assume that the two batteries at the relay have the same initial state. In each block, the battery \#1 (battery \#2) duplicates in the even (odd) slot the operational mode that is operated by the battery \#2 (battery \#1) in the odd (even) slot, so that the two batteries have the same energy state at the beginning (end) of each block.
Based on the considered discretized battery model, we define the energy that can be harvested from the received signal to be equal to $\varepsilon \triangleq \varepsilon _{i_h^*}$, where
\begin{eqnarray}
i_h^* &\!\!=\!\!& \arg \mathop{\max}\limits_{i\in \{ 0, \ldots, L+1 \}} \left\{ \varepsilon_i:  \begin{array}{*{20}{c}}
{}\\
{}
\end{array} \right. \nonumber \\
&\!\!\!\!  & \left. \varepsilon _i <  \left\{ {\begin{array}{*{3}{c}}
 \eta p_s g_1, & \mu(t) = \mu_h \\
 \eta \rho (p_s g_1+p_r g_a) , & \mu(t) = \mu_{hr}
\end{array}} \right.  \right\}. \label{eq:epsilon_h}
\end{eqnarray}
As for the relay transmission, the relay also uses the $L+2$ discrete energy levels. Corresponding to $p_r$ of \eqref{eq:p_r_acc}, the required transmitted energy level is given by
\begin{eqnarray}
\varepsilon'\triangleq \left\{ {\begin{array}{*{20}{c}}
\varepsilon_{i_r^*},  & {\rm  if~~} p_r \le p_b  \\
\infty,  &  {\rm  otherwise}
\end{array}} \right.,
\end{eqnarray}
where $i_r^* = \arg \mathop{\min}\limits_{i\in \{ 1, \ldots, L+1 \}}\{ \varepsilon_i: \varepsilon_i \ge p_r \} $.

Since an outage event occurs when the source signal cannot be decoded at the destination or equivalently when the relay operates in the mode of $\mu_{\phi}$ or $\mu_{h}$, the main optimization target is to minimize the number of times that the relay  does not transmit. To this end, the GS policy prioritizes the operation modes $\mu_r$ and $\mu_{hr}$. When the residual energy in the two batteries can support the required transmitted energy, the GS policy switch the relay to transmission, otherwise it switch the relay to EH. Further, when the relay is allowed to transmit, the GS policy prefers $\mu_{hr}$ more than $\mu_r$ in order to harvest the self-emitted energy whenever it is possible.  For the $t$-th block, the GS policy can be expressed as
\begin{eqnarray}
\mu^{(\rm{GS})}(t) =  \left\{ {\begin{array}{*{20}{c}}
\mu_{hr},   & E_0(t-1) \ge \varepsilon' {\rm ~~and~~} \varepsilon \ge \varepsilon_1  \\
\mu_r   ,   & E_0(t-1) \ge \varepsilon' {\rm ~~and~~} \varepsilon < \varepsilon_1    \\
\mu_h   ,   & E_0(t-1) < \varepsilon'   {\rm ~~and~~} \varepsilon \ge \varepsilon_1  \\
\mu_{\phi}, & {\rm otherwise}
\end{array}} \right. \label{eq:GS_states}
\end{eqnarray}
and
\begin{eqnarray}
E_0(t) \!=\! \min\{p_b,  E_0(t\!-\!1) - w_1(t\!-\!1) \varepsilon'  + w_2(t\!-\!1) \varepsilon   \},
\end{eqnarray}
where $w_1(t) \triangleq \mathbb{I}\{\mu(t) = \mu _r, \mu_{hr} \}$ and $w_2(t) \triangleq \mathbb{I}\{\mu(2t) = \mu _h, \mu_{hr}  \}$ are the binary variables and $\mathbb{I}$ denotes the indicator function.

By employing the GS policy, the relay's battery group transits among the harvesting, relaying, harvesting-relaying, and idle behaviors, which can be represented by a finite Markov chain. Due to the complicated forms of the considered fading distribution, it is hard to derive the outage probability in a closed-form and we use simulations to verify the outage performance of the GS policy.

\section{Simulation Results}

\begin{figure}[htbp]
\begin{center}
\includegraphics[width=2.4in]{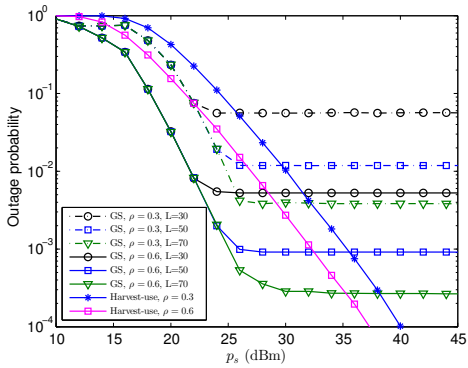}
\vspace{-0.09in}
\caption{Outage probability versus $p_s$.}
\end{center}
\vspace{-0.25in}
\end{figure}

In this section, we consider a single carrier system working with the carrier frequency 868 MHz and a bandwidth 200 kHz. The required SINR threshold at the destination for decoding is set as $\gamma_{\rm th} = 7$, which corresponds to a fixed transmission rate of 3 bps/Hz. The distances between the source and relay and between the relay and destination are set as $d_1=8$ m and $d_2=18$ m, respectively. The large-scale path coefficients are set as $\mathcal{L}_1=3.49 \times 10^{-4}$ and $\mathcal{L}_2=4.59 \times 10^{-6}$, respectively. The noise powers are set by $\sigma_a^2 = -100$ dBm, $\sigma_p^2 = -90$ dBm, $\sigma_d^2 = -90$ dBm, respectively.
The energy conversion efficiency is set as $\eta=0.3$, and the EH receiver sensitivity is set as $\varepsilon_{\min} =-27$ dBm \cite{CMOS_Rectifier}.
The size of each battery is set as $p_b = m_1 \theta_1 p_s$. Considering $\varepsilon_{\min}$ in practice, we define the number of the energy levels of each battery by $\tilde L +2$, where $\tilde L \triangleq \min\{L,  \lfloor p_b/\varepsilon_{\min}  \rfloor \}$. Notably, we have considered both the EH outage and information outage in the evaluation.

In Fig. 2, the outage probability versus $p_s$ is investigated for the virtual harvest-transmit model and harvest-transmit-store model (with its GS implementation). In the simulation, we set $m_1=4$, $m_2=2$, $g_a=-10$ dB, and $g_b=-20$ dB. As can be seen, the GS policy achieves the smallest outage probability in the low and middle $p_s$ regions. Further,
Fig. 2 shows that the number of energy level of the battery group is a critical parameter for the performance of the GS policy. In the low and middle $p_s$ regions, it can be shown that the three different $L$s result in the same outage probability for the GS policy. Nevertheless, in the high $p_s$ region, the outage probability achieved by the GS policy decreases with the increasing $L$.
Additionally, the GS policy suffers from an outage floor, so that the GS policy achieves zero diversity order in the high $p_s$ region. In summary, Fig. 2 verifies that the energy scheduling is superior to the virtual harvest-transmit model in the low $p_s$ region.
Notably, in Fig. 2, the outage floor corresponding to PS-SWIPT with the virtual harvest-transmit model does not appear due to the significantly reduced RSI.

\section{Conclusion}

This paper has investigated PS-SWIPT in an AF FDR network. The virtual harvest-transmit model and harvest-transmit-store model have been proposed. With the aid of time-switched battery group for charging and discharging, the concurrent source and relay transmissions have been enabled by a PS-operated FDR node. To effectively use the harvested energy, a GS policy has been designed for the harvest-transmit-store model. Simulation results verify the outage performance of the proposed schemes. It has shown the harvest-transmit-store model is superior to the virtual harvest-transmit model in the low source transmission power region.

\bibliography{IEEEabrv,IEEE_bib}

\begin{thebibliography}{10}
\providecommand{\url}[1]{#1}
\csname url@samestyle\endcsname
\providecommand{\newblock}{\relax}
\providecommand{\bibinfo}[2]{#2}
\providecommand{\BIBentrySTDinterwordspacing}{\spaceskip=0pt\relax}
\providecommand{\BIBentryALTinterwordstretchfactor}{4}
\providecommand{\BIBentryALTinterwordspacing}{\spaceskip=\fontdimen2\font plus
\BIBentryALTinterwordstretchfactor\fontdimen3\font minus
  \fontdimen4\font\relax}
\providecommand{\BIBforeignlanguage}[2]{{%
\expandafter\ifx\csname l@#1\endcsname\relax
\typeout{** WARNING: IEEEtran.bst: No hyphenation pattern has been}%
\typeout{** loaded for the language `#1'. Using the pattern for}%
\typeout{** the default language instead.}%
\else
\language=\csname l@#1\endcsname
\fi
#2}}
\providecommand{\BIBdecl}{\relax}
\BIBdecl

\bibitem{Energy_harvesting_relay}
C.~Huang, R.~Zhang, and S.~Cui, ``Throughput maximization for the gaussian
  relay channel with energy harvesting constraints,'' \emph{IEEE J. Sel. Areas
  in Commun.}, vol.~31, no.~8, pp. 1469--1479, Aug. 2013.

\bibitem{SWIPT_protocol_AF}
A.~Nasir, X.~Zhou, S.~Durrani, and R.~Kennedy, ``Relaying protocols for
  wireless energy harvesting and information processing,'' \emph{IEEE Trans.
  Wireless Commun.}, vol.~12, no.~7, pp. 3622--3636, Jul. 2013.

\bibitem{Grover_Shannon_meets_Tesla}
P.~Grover and A.~Sahai, ``Shannon meets {Tesla}: Wireless information and power
  transfer,'' in \emph{Proc. IEEE Int. Symp. Inf. Theory}, Austin, TX, Jun.
  2010, pp. 2363--2367.

\bibitem{SWIPT_cellular}
K.~Huang and V.~Lau, ``Enabling wireless power transfer in cellular networks:
  Architecture, modeling and deployment,'' \emph{IEEE Trans. Wireless Commun.},
  vol.~13, no.~2, pp. 902--912, Feb. 2014.

\bibitem{SWIPT_magazine}
I.~Krikidis, S.~Timotheou, S.~Nikolaou, G.~Zheng, D.~Ng, and R.~Schober,
  ``Simultaneous wireless information and power transfer in modern
  communication systems,'' \emph{IEEE Commun. Mag., Green Communications and
  Computing Networks Series}, vol.~52, no.~11, pp. 104--110, Nov. 2014.

\bibitem{SWIPT_architecture}
X.~Zhou, R.~Zhang, and C.~K. Ho, ``Wireless information and power transfer:
  Architecture design and rate-energy tradeoff,'' \emph{IEEE Trans. Commun.},
  vol.~61, no.~11, pp. 4754--4767, Nov. 2013.

\bibitem{SWIPT_OFDM}
D.~W.~K. Ng, E.~S. Lo, and R.~Schober, ``Energy-efficient resource allocation
  in multiuser {OFDM} systems with wireless information and power transfer,''
  in \emph{Proc. IEEE WCNC 2013}, Shanghai, China, April 2013, pp. 3823--3828.

\bibitem{MIMO_B_SWIPT}
R.~Zhang and C.~K. Ho, ``{MIMO} broadcasting for simultaneous wireless
  information and power transfer,'' \emph{IEEE Trans. Wireless Commun.},
  vol.~12, no.~5, pp. 1989--2001, May 2013.

\bibitem{SWIPT_SRR}
Z.~Ding, I.~Krikidis, B.~Sharif, and H.~V. Poor, ``Wireless information and
  power transfer in cooperative networks with spatially random relays,''
  \emph{IEEE Trans. Wireless Commun.}, vol.~13, no.~8, pp. 4440--4453, Aug.
  2014.

\bibitem{SWIPT_DPS}
H.~Chen, Y.~Li, Y.~Jiang, Y.~Ma, and B.~Vucetic, ``Distributed power splitting
  for {SWIPT} in relay interference channels using game theory,'' \emph{IEEE
  Trans. Wireless Commun.}, vol.~14, no.~1, pp. 410--420, Aug. 2014.

\bibitem{SWIPT_PA}
Z.~Ding, S.~M. Perlaza, I.~Esnaola, and H.~V. Poor, ``Power allocation
  strategies in energy harvesting wireless cooperative networks,'' \emph{IEEE
  Trans. Wireless Commun.}, vol.~13, no.~2, pp. 846--860, Feb. 2014.

\bibitem{SWIPT_antenna_switch}
I.~Krikidis, S.~Sasaki, S.~Timotheou, and Z.~Ding, ``A low complexity antenna
  switching for joint wireless information and energy transfer in {MIMO} relay
  channels,'' \emph{IEEE Trans. Commun.}, vol.~62, no.~5, pp. 1577--1587, May
  2014.

\bibitem{SWIPT_Antenna_select}
Z.~Zhou, M.~Peng, Z.~Zhao, and Y.~Li, ``Joint power splitting and antenna
  selection in energy harvesting relay channels,'' \emph{IEEE Trans. Signal
  Process.}, vol.~22, no.~7, pp. 823--827, Jul. 2015.

\bibitem{SWIPT_AF_DPS2}
L.~Hu, C.~Zhang, and Z.~Ding, ``Dynamic power splitting policies for {AF} relay
  networks with wireless energy harvesting,'' in \emph{Proc. IEEE ICC 2015},
  London, UK, 8-12, June 2015, pp. 1--5.

\bibitem{SWIPT_FD_selfenergy}
Y.~Zeng and R.~Zhang, ``Full-duplex wireless-powered relay with self-energy
  recycling,'' \emph{IEEE Wireless Commun. Lett.}, vol.~4, no.~2, pp. 201--204,
  2015.

\bibitem{SWIPT_FD}
C.~Zhong, H.~Suraweera, G.~Zheng, I.~Krikidis, and Z.~Zhang, ``Wireless
  information and power transfer with full duplex relaying,'' \emph{IEEE Trans.
  Commun.}, vol.~62, no.~10, pp. 3447--3461, Oct. 2014.

\bibitem{SWIPT_FDR_MIMO}
M.~Mohammadi, H.~Suraweera, G.~Zheng, C.~Zhong, and I.~Krikidis, ``Full-duplex
  {MIMO} relaying powered by wireless energy transfer,'' in \emph{Proc. 16th
  International Workshop on Signal Processing Advances in Wireless
  Communications (SPAWC)}, Stockholm, Sweden, Jun. 2015, pp. 296--300.

\bibitem{Optimal_Training_WET}
Y.~Zeng and R.~Zhang, ``Optimized training design for wireless energy
  transfer,'' \emph{IEEE Trans. Commun.}, vol.~63, no.~2, pp. 536--550, 2015.

\bibitem{Lifetime_max}
W.~J. Huang, Y.~W.~P. Hong, and C.~C.~J. Kuo, ``Lifetime maximization for
  amplify-and-forward cooperative networks,'' \emph{IEEE Trans. Wireless
  Commun.}, vol.~7, no.~5, pp. 1800--1805, 2008.

\bibitem{Relaying_vs_EH}
I.~Krikidis, S.~Timotheou, and S.~Sasaki, ``{RF} energy transfer for
  cooperative networks: Data relaying or energy harvesting?'' \emph{IEEE
  Commun. Letters}, vol.~16, no.~11, pp. 1772--1775, 2012.

\bibitem{CMOS_Rectifier}
M.~Stoopman, S.~Keyrouz, H.~J. Visser, K.~Philips, and W.~A. Serdijn,
  ``Co-design of a {CMOS} rectifier and small loop antenna for highly sensitive
  {RF} energy harvesters,'' \emph{IEEE J. Solid-State Circuits}, vol.~49,
  no.~3, pp. 622--634, 2014.

\end{thebibliography}

\balance

\end{document}